\documentclass[aps,prl,twocolumn,superscriptaddress,showpacs,amsmath,amssymb]{revtex4-2}

\usepackage{graphicx}
\usepackage{bm}
\usepackage{dcolumn}
\usepackage[colorlinks=true,linkcolor=blue,citecolor=blue,urlcolor=blue]{hyperref}

\newcommand{\nhat}{\hat{\bm{m}}}
\newcommand{\R}{\bm{R}}
\newcommand{\EYSR}{E_{\mathrm{YSR}}}

\begin{document}

\title{Driving Atomic Spins through Their Substrate Electrons}

\author{Samir Lounis}
\affiliation{Institute of Physics and Halle-Berlin-Regensburg Cluster of Excellence CCE, Martin-Luther-University Halle-Wittenberg, 06099 Halle (Saale), Germany}

\date{\today}

\begin{abstract}
Coherent control of single spins on surfaces has relied on decoupling them from the substrate with insulating films, which protects the spin but also removes any electronic means of addressing it. Here we show that a local electric potential can drive an adatom spin through the substrate electrons themselves. An exact reciprocity relation fixes the torque, which equals the change of the local electron density when the moment rotates; the drive map is thus a ground-state property that can be measured independently. The relation also implies sharp selection rules, since point-contact spins, spin-$1/2$ and the central transition of half-integer spins remain dark. On a normal metal the hybridization responsible for the drive also destroys coherence, but on a superconductor the drive survives while substrate damping is confined to subgap transitions. For atomically sharp tips we find Rabi frequencies of several MHz at RF amplitudes of tens of mV, with substrate-limited linewidths below $0.1$~MHz, so that strong hybridization, usually the enemy of spin coherence, becomes an asset.
\end{abstract}

\maketitle

\textit{Introduction.}---Electron spin resonance in a scanning tunneling microscope (ESR-STM) has turned single atoms on surfaces into addressable quantum objects~\cite{Baumann2015,Chen2022}. The price of this achievement is decoupling. The magnetic atom sits on a thin insulating film, typically MgO on Ag, so that substrate electrons cannot flip its spin within picoseconds. The same film also cuts the electronic channels through which the spin could be controlled, so the drive is borrowed from elsewhere. With an antenna, the transitions can be driven by the ac magnetic field of the radiation itself~\cite{Ast2024}. Otherwise, in the prevailing picture, the RF field displaces the atom in the inhomogeneous exchange or magnetic field of a spin-polarized tip~\cite{Lado2017,Seifert2020SA,Chen2022}, with transverse field gradients driving spin-$1/2$ and longitudinal ones driving integer spins~\cite{Seifert2020SA}; cotunnelling~\cite{ReinaGalvez2019,ReinaGalvez2025}, spin-transfer torque~\cite{Kovarik2024}, modulation of the $g$-factor anisotropy~\cite{Ferron2019}, electric-field-induced anisotropy~\cite{Sonntag2014} and driving through a neighbouring magnetic atom~\cite{Phark2023} have also been discussed~\cite{Delgado2021}, and a bias-dependent spin-electric shift of the resonance has been observed~\cite{Kot2023}. All of these act through the tip, the junction or a second spin. None uses the substrate electrons that surround the atom: the mechanism proposed here needs neither a magnetic tip nor a second spin, but requires spin-orbit coupling and spatially extended hybridization.

On a bare metal the situation is reversed: hybridization gives the substrate electrons a strong grip on the spin, but electrons at the Fermi level relax it almost immediately~\cite{Korringa1950,Lounis2010,Khajetoorians2011,Khajetoorians2013,Schweflinghaus2016}. The same coupling shapes the orientation of the spin through its local electronic environment~\cite{Noei2023}, and can even be strong enough to dress the substrate electrons with spin excitations~\cite{Schweflinghaus2014}, forming spinarons~\cite{Bouaziz2020,Friedrich2024}. Even on MgO, scattering of tunnelling and substrate electrons sets the measured coherence times~\cite{Willke2018}, which is why decoupling has become the guiding principle. Strong coupling and coherence seem to exclude each other.

Here we show that the environment one usually isolates the spin from can provide the drive. The substrate electrons do two different jobs. Their \emph{anisotropic charge response} to a local potential drives the spin; their \emph{low-energy excitations} damp it. The first is a property of the whole Fermi sea, the second of states within a few $k_BT$ of the Fermi level. A superconductor removes the second and leaves the first untouched. Gap protection of excited magnetic states is known~\cite{Heinrich2013}, spin excitations of adatom assemblies on superconductors have been resolved~\cite{Brinker2022}, and Yu-Shiba-Rusinov (YSR) states~\cite{Yu1965,Shiba1968,Rusinov1969,Balatsky2006} have been proposed as qubits~\cite{Mishra2021} and excited by microwaves~\cite{MicrowaveYSR2023,Ismail2025}. What has not been recognized is that strong hybridization can be \emph{exploited} for electrical driving once the gap removes its cost.

\textit{An exact identity.}---Consider a magnetic adatom with moment direction $\nhat$, coupled to substrate electrons through orbital-selective hybridization, exchange and spin-orbit coupling (SOC), and a local potential $v(t)$ at the tip position $\R$. For the grand potential $\Omega(\nhat,v)$, the Hellmann-Feynman theorem gives $\partial\Omega/\partial v=n(\R;\nhat)$, the electron density at the tip position, while $\partial\Omega/\partial\nhat$ is the exchange field acting on the moment. Equality of the mixed derivatives, a Maxwell-type relation, yields the torque exerted by the potential,
\begin{equation}
\bm{\tau}(\R)=v\,\frac{\partial n(\R;\nhat)}{\partial\nhat}.
\label{eq:identity}
\end{equation}
Here $\bm\tau$ is the generalized torque, conjugate to the angular coordinates of $\nhat$: $\partial/\partial\nhat$ is the derivative on the unit sphere, with components $\tau_\theta=v\,\partial n/\partial\theta$ and $\tau_\varphi=v\,(\sin\theta)^{-1}\partial n/\partial\varphi$; the mechanical torque vector follows as $\nhat\times\bm\tau$ up to sign. The Maxwell relation itself holds at any $v$; Eq.~(\ref{eq:identity}) is its linear-response form~\cite{SM}. The electrically driven torque is exactly the anisotropic part of the charge density around the adatom. Reciprocity of this kind underlies the electric-field control of magnetic anisotropy at metal surfaces, where spin-dependent screening links the field-induced anisotropy to the induced charge~\cite{Duan2008,Nakamura2009,Nakamura2009b,Brovko2014}; here it applies to a local potential and makes the ESR drive a measurable property of the local density. We verified Eq.~(\ref{eq:identity}) to four digits against a direct linear-response calculation~\cite{SM}.

\begin{figure}[t]
\includegraphics[width=\columnwidth]{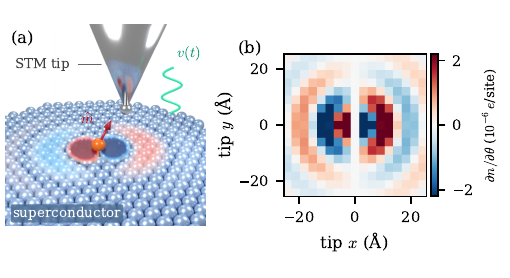}
\caption{(a) Mechanism: a local potential $v(t)$ under the tip reshapes the electron density around the adatom; because the density depends on the moment direction, the potential exerts a torque. (b) Drive map $\partial n(\R;\nhat)/\partial\theta$ versus tip position for a moment tilted from the surface normal towards $x$ (Au(111)-like model, atomic SOC $\xi=60$~meV). By Eq.~(\ref{eq:identity}) it is both the electric torque per unit potential and the density change measured by STM when the moment is rotated.}
\label{fig:mechanism}
\end{figure}

The derivation is elementary; its value lies in what it makes accessible. The drive map becomes a ground-state property, computable with any equilibrium method, including first-principles ones. Above all, it becomes \emph{measurable without ESR}. Rotating the moment with an external field changes the local density of states $\rho(\R,E)$, as observed in anisotropic-magnetoresistance STM~\cite{Wiesendanger2009}; the drive map is the energy integral of this change over occupied states, $\partial n/\partial\nhat=\int^{E_F}\!dE\,\partial\rho/\partial\nhat$. Field-rotation spectroscopic maps at negative bias therefore reconstruct it, and its spatial pattern, which does not require the full energy integral, provides the sharpest test. No other driving mechanism predicts this correspondence. Read in reverse, Eq.~(\ref{eq:identity}) also describes the density modulation produced by a precessing moment, the spin-orbit-mediated ``breathing'' once proposed as an ESR detection channel~\cite{Balatsky2012}: drive and signal are reciprocal partners. For a moment tilting away from the surface normal the map is dipolar, vanishes with the tip directly above the atom and peaks at offsets comparable to the atomic radius [Fig.~\ref{fig:mechanism}(b)].

\textit{Selection rules.}---For drives acting through time-reversal-even charge redistribution, at linear order in $v$, the identity constrains which spins can be driven. First, \emph{point-contact exchange gives no drive}: in a time-reversal-symmetric host the on-site Green function is proportional to the unit matrix in spin space, so an adatom coupled through a single contact exchange is rotationally invariant, and the tip potential changes only the magnitude of the induced polarization, never its direction. The drive measures the spatial and orbital extent of the coupling.

\begin{figure}[t]
\includegraphics[width=\columnwidth]{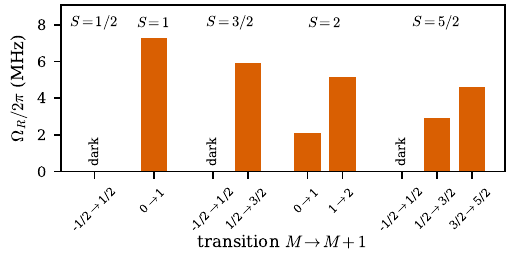}
\caption{Rabi frequencies of the $\Delta M=+1$ transitions for $S=1/2$ to $5/2$ from Eq.~(\ref{eq:quantum}), for an RF amplitude of $30$~mV and a tip of $1$~nm radius at a $6$~\AA\ gap~\cite{SM}. Spin-$1/2$ and the central transition of half-integer spins are dark.}
\label{fig:selection}
\end{figure}

Second, time reversal requires $n(\R;\nhat)=n(\R;-\nhat)$, so the drive is a modulated anisotropy, $\delta E=v\,\nhat\cdot Q(\R)\cdot\nhat$. Quantizing this classical anisotropy through the spin-coherent-state correspondence, a semiclassical step that we test below by exact diagonalization, gives for $v(t)=v\cos\omega t$
\begin{equation}
\delta H=\frac{v\,Q_{xz}}{S(S-\tfrac12)}\,\{S_x,S_z\}+\dots,\qquad \tau=2vQ_{xz}.
\label{eq:quantum}
\end{equation}
For $S=1/2$ every quadratic spin operator is a constant: \emph{spin-$1/2$ is not driven}. The $M\to M+1$ matrix element carries a factor $(2M+1)$: \emph{the central transition of half-integer spins is dark} (Fig.~\ref{fig:selection}). These rules distinguish the mechanism from tip-based ones, which drive spin-$1/2$ efficiently. A nonmagnetic impurity is no substitute: time reversal then forbids a static spin response, and the dynamical one~\cite{Etxeberria2026} is suppressed as $\hbar\omega/E_F\sim10^{-3}$--$10^{-4}$~\cite{SM}.

\textit{Drive law.}---To make the physics concrete we use a two-dimensional Rashba electron gas with Au(111) surface-state parameters and a $d$-orbital adatom (exchange splitting $2.5$~eV, atomic SOC $\xi$), whose orbital-selective hybridization with the surface is scaled by a factor $s$~\cite{SM}. The model serves to expose how drive and damping depend on coupling, not to describe a specific material. The electric potential is that of a spherical tip of radius $R_t$ at a $6$~\AA\ gap; because only its laterally asymmetric part produces a torque, sharp tips are favoured~\cite{SM}.

\begin{figure}[t]
\includegraphics[width=\columnwidth]{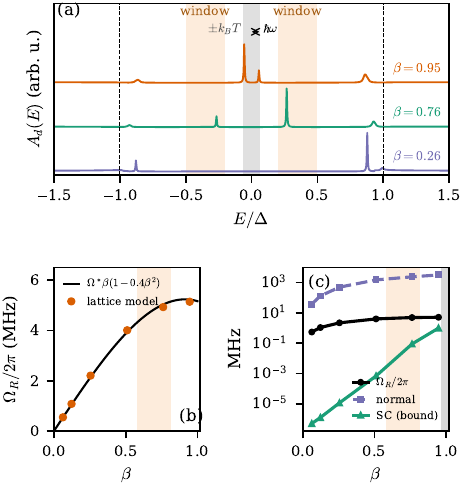}
\caption{(a) Adatom spectral function in the gap for three pair-breaking strengths $\beta$ (curves offset): the YSR peak moves from the gap edge (dashed) towards the Fermi level. Shaded: the operating window $\EYSR\approx0.2$--$0.5\Delta$ and the thermal band $\pm k_BT$; the ESR quantum $\hbar\omega$ is shown for scale. (b) Rabi frequency ($S=2$, $1\to2$; $30$~mV RF, $R_t=1$~nm) versus $\beta$: lattice model (dots) and Eq.~(\ref{eq:drivelaw}) with $c=0.4$ (line). (c) Rabi frequency and substrate-limited linewidth at $20$~GHz on a normal metal and on a superconductor ($\hbar\omega/\Delta=0.06$, $k_BT/\Delta=0.065$); superconducting values are upper bounds. Orange shading in (b,c): the operating window $\EYSR\approx0.2$--$0.5\Delta$; grey: the parity crossing.}
\label{fig:window}
\end{figure}

The torque and the pair-breaking strength of the adatom are both of second order in the hybridization. On a superconductor the latter is the YSR parameter $\beta$: as the coupling grows, the YSR peak moves from the gap edge towards the Fermi level [Fig.~\ref{fig:window}(a)]. We define $\beta$ operationally through the lowest YSR energy, using Shiba's classical relation~\cite{Shiba1968},
\begin{equation}
\frac{\EYSR}{\Delta}=\frac{1-\beta^2}{1+\beta^2}.
\label{eq:shiba}
\end{equation}
Scaling all hybridization amplitudes by a factor $s$ ($s=1$ for the reference parameters), we find $\beta\simeq s^2$, as expected for a quantity of second order in the hybridization. Since the torque is of the same order, the drive is linear in $\beta$ at weak coupling. It saturates when the hybridization width $\Gamma$ of the $d$ levels becomes comparable to an energy $\varepsilon^*$ of the order of their distance from the Fermi level: varying $\Gamma$ at fixed $s$ gives $\tau\propto[1+(\Gamma/\varepsilon^*)^2]^{-1}$ with $\varepsilon^*\approx0.46$~eV~\cite{SM}. Because $\Gamma\propto\beta$, this yields
\begin{equation}
\Omega_R\simeq\Omega^*\,\beta\left(1-c\,\beta^2\right),\qquad c=\big[\Gamma(\beta{=}1)/\varepsilon^*\big]^2\approx0.43,
\label{eq:drivelaw}
\end{equation}
which reproduces the lattice results to within a few percent up to the parity crossing [Fig.~\ref{fig:window}(b)]; a direct fit gives $c=0.4$. The prefactor $\Omega^*$ is the torque per unit $\beta$ at weak coupling. For $S=2$, an RF amplitude of $30$~mV and $R_t=1$~nm, $\Omega^*/2\pi\approx9$~MHz; it scales linearly with the RF amplitude and is three times smaller for $R_t=3$~nm. The drive is dominated by the adatom's own SOC and changes by only $3$--$4\%$ when the substrate turns superconducting~\cite{SM}.

\textit{Damping.}---On a normal metal the same hybridization opens fast relaxation. The Gilbert damping from the torque-torque correlation at the Fermi level~\cite{Kambersky1976} grows faster with coupling than the drive, and the linewidth exceeds the Rabi frequency by two to three orders of magnitude over the whole range [Fig.~\ref{fig:window}(c)].

Generally, the damping follows from the dissipative part of the torque-torque correlation function at the ESR frequency, the finite-frequency form of Kamberský's torque-correlation result~\cite{Kambersky1976,Kambersky2007,Guimaraes2019,SM}, equivalent to the damping extracted from dynamical transverse susceptibilities of adatoms~\cite{Lounis2010,Lounis2015}. On a superconductor with $\hbar\omega,k_BT\ll\Delta$, it reduces to a golden-rule sum over subgap Bogoliubov levels $a,b$,
\begin{equation}
\alpha(\omega)=\frac{\pi}{2S\omega}\sum_{a,b}|T_{ab}|^2\,[f_a-f_b]\,\delta(\hbar\omega-E_b+E_a),
\label{eq:goldenrule}
\end{equation}
with $T_{ab}$ the matrix elements of the torque operator in the Nambu basis; the factor $1/2$ removes the Bogoliubov-de Gennes double counting~\cite{SM}. Three channels contribute, each yielding a design rule. (i) Intra-level fluctuations, with $T_{aa}=\partial E_a/\partial\theta$, vanish to first order when the quantization axis is a symmetry axis of the adsorption site, where $\EYSR(\theta)$ is even. (ii) Quasiparticle transfer between YSR levels is suppressed as $e^{-E_{\min}/k_BT}$ when the lowest YSR state lies several $k_BT$ above the Fermi level. (iii) Two-quasiparticle creation conserves parity and survives at $T=0$, but only if $\hbar\omega$ exceeds the lowest pair energy $E_a+E_b$.

When these conditions hold, the substrate no longer limits coherence. For $\hbar\omega/\Delta=0.06$ and $k_BT/\Delta=0.065$ (20~GHz and 1~K for Pb), the damping drops by three to eight orders of magnitude relative to the normal metal [Fig.~\ref{fig:window}(c)]; the computed values are upper bounds set by the phenomenological YSR broadening~\cite{SM}. At $\EYSR\approx0.27\Delta$ the substrate-limited linewidth is below $0.1$~MHz while the drive is about $5$~MHz. The results depend only on dimensionless ratios and transfer to the real Pb gap~\cite{SM}.

\textit{Quantum validation.}---The classical-moment treatment is weakest near the parity crossing, and Eq.~(\ref{eq:quantum}) was so far inferred. We therefore solve a quantum problem in which the drive emerges microscopically: a spin $S=2$ with easy-axis anisotropy is exchange-coupled to two sites of a four-site square plaquette superconducting $2\times2$ cluster with Rashba SOC $\lambda$, the tip potential acts on a third site, and the drive is the matrix element of the tip-site density between ESR states, obtained by exact diagonalization in each parity sector~\cite{SM}.

\begin{figure}[t]
\includegraphics[width=\columnwidth]{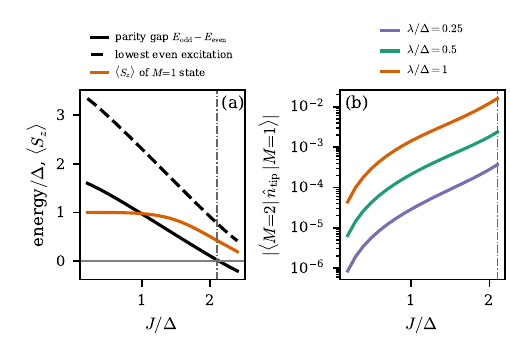}
\caption{Quantum spin $S=2$ exchange-coupled to a four-site square plaquette superconducting $2\times2$ Rashba cluster. (a) Parity gap, lowest parity-conserving excitation and $\langle S_z\rangle$ of the $M=1$ ESR state versus exchange $J$ ($\lambda/\Delta=0.5$). (b) Microscopic drive, the matrix element of the tip-site density between the $M=1$ and $M=2$ states, for three SOC strengths, up to the parity crossing (vertical line).}
\label{fig:quantum}
\end{figure}

This microscopic drive reproduces every selection rule. It vanishes without SOC and for point-contact coupling. The ratio of its $1\to2$ and $0\to1$ elements is $2.50$, against $2.45$ from Eq.~(\ref{eq:quantum}). For $S=3/2$ the central transition is suppressed $550$-fold, and for $S=1/2$ the drive appears only in proportion to the Zeeman field, beyond linear time-reversal-even order. The drive grows steeply with exchange, roughly as $J^{2.5}$, as the parity gap closes [Fig.~\ref{fig:quantum}(b)]. Close to the crossing, however, the ESR multiplet distorts once the parity gap falls below about $0.5\Delta$ [Fig.~\ref{fig:quantum}(a)]. The operating window is therefore $\EYSR\approx0.2$--$0.5\Delta$: close enough to the Fermi level for a strong drive, far enough for an intact multiplet.

Parity also strengthens the protection. Mean field treats the YSR occupation as thermally fluctuating, which underlies channels (i) and (ii). In the quantum problem, entering the odd sector requires a quasiparticle from outside, while the lowest parity-conserving excitation stays far above $\hbar\omega$ [Fig.~\ref{fig:quantum}(a)]. Coherence near the crossing is then limited by quasiparticle poisoning, a measurable property of the superconductor, rather than by thermal excitation.

\textit{Feasibility.}---Since $\hbar\omega$ lies far below all electronic energy scales, the electrons follow the RF potential adiabatically, and a Floquet description, as used for all-electric ESR~\cite{ReinaGalvez2021}, reduces to the static identity, Eq.~(\ref{eq:identity}), and the linear-response damping, Eq.~(\ref{eq:goldenrule}). The RF potential also modulates $\EYSR$ during each cycle, but only weakly: a $5$~meV surface potential shifts it by about $2\times10^{-4}\Delta$, and an additional $10$~meV shift of the adatom levels by about $5\times10^{-3}\Delta$~\cite{SM}, far from driving the bound state through zero. The coherence budget then has three distinct parts: the intrinsic substrate damping computed above, parity switching by quasiparticle poisoning, and decoherence induced by the junction. Once the substrate is no longer the bottleneck, the junction dominates. For comparison, antenna-driven experiments reach RF amplitudes of $\approx20$~mV at the junction and Rabi frequencies of $\approx7$~MHz~\cite{Ast2024}, comparable to the drive found here. Tunnelling electrons relax the spin at a rate of order $pI/e$~\cite{Ast2024}; at $I=10$~pA and a spin-flip probability $p\approx0.05$ this is about $0.5$~MHz. Since the drive needs a field rather than a current, the tip can be retracted during driving, and RF delivery through an antenna rather than the bias line~\cite{Seifert2020PRR} decouples the drive from the junction further: $3$~\AA\ reduces the current about four-hundred-fold, while the voltage needed for the same field rises by only $40\%$. A pulsed protocol, with driving at large tip height and readout at small height, then keeps junction decoherence well below the drive. Suitable systems combine an integer spin $S\geq1$, a YSR state at $0.2$--$0.5\Delta$ and an ESR transition well below $2\Delta$. Mn, Cr and Fe adatoms on Pb(111), whose YSR spectra have been mapped in detail and depend on adsorption site and spin-orbit coupling~\cite{Yazdani1997,Ji2008,Ruby2016,Choi2017,Heinrich2018,Cornils2017,Beck2021}, are natural candidates. Nb(110), on which atom-by-atom assembled spin chains~\cite{Kuester2021a,Schneider2021} and their Josephson response~\cite{Kuester2021b} have been studied, and NbSe$_2$ offer larger gaps than Pb, while V(100) is already used for microwave STM~\cite{MicrowaveYSR2023}. Crucially, the coupling of magnetic adsorbates to a superconductor can be tuned across the parity crossing~\cite{Franke2011,Farinacci2018,Kuester2021a}, so the YSR energy can be placed inside the operating window.

\textit{Conclusion.}---The trade-off between electrical control and coherence of surface spins is not fundamental. An exact identity ties the electric drive to the anisotropic charge density, making the drive map a measurable ground-state property and implying selection rules that set the mechanism apart. On normal metals the hybridization that creates the drive destroys coherence; on superconductors it does not. The broader message is a change of design principle: rather than decoupling spins from their substrate, couple them strongly to a substrate whose low-energy excitations are gapped, and use the YSR energy as the knob that balances drive and protection.

\begin{acknowledgments}
Funding has been provided by the Deutsche Forschungsgemeinschaft (DFG)  through
Project No.~328545488 -- CRC/TRR 227, Project No.~B12, and Projects LO 1659/10-1, LO 1659/11-1.
\end{acknowledgments}



\begin{thebibliography}{99}

\bibitem{Baumann2015} S. Baumann, W. Paul, T. Choi, C. P. Lutz, A. Ardavan, and A. J. Heinrich, Electron paramagnetic resonance of individual atoms on a surface, Science \textbf{350}, 417 (2015).

\bibitem{Chen2022} Y. Chen, Y. Bae, and A. J. Heinrich, Harnessing the quantum behavior of spins on surfaces, Adv. Mater. \textbf{35}, 2107534 (2023). 

\bibitem{Lado2017} J. L. Lado, A. Ferr\'on, and J. Fern\'andez-Rossier, Exchange mechanism for electron paramagnetic resonance of individual adatoms, Phys. Rev. B \textbf{96}, 205420 (2017).

\bibitem{ReinaGalvez2019} J. Reina G\'alvez, C. Wolf, F. Delgado, and N. Lorente, Cotunneling mechanism for all-electrical electron spin resonance of single adsorbed atoms, Phys. Rev. B \textbf{100}, 035411 (2019).

\bibitem{ReinaGalvez2025} J. Reina-G\'alvez \textit{et al.}, Contrasting exchange-field and spin-transfer torque driving mechanisms in all-electric electron spin resonance, Phys. Rev. B \textbf{112}, 245408 (2025); Erratum (2026). 

\bibitem{Phark2023} S.-h. Phark \textit{et al.}, Electric-field-driven spin resonance by on-surface exchange coupling to a single-atom magnet, Adv. Sci. (2023). 

\bibitem{Ast2024} C. R. Ast, P. Kot, M. Ismail, S. de-la-Pe\~na, A. I. Fern\'andez-Dom\'inguez, and J. C. Cuevas, Theory of electron spin resonance in scanning tunneling microscopy, arXiv:2403.20247 (2024). 

\bibitem{Delgado2021} F. Delgado and N. Lorente, A theoretical review on the single-impurity electron spin resonance on surfaces, Prog. Surf. Sci. \textbf{96}, 100625 (2021).

\bibitem{Kot2023} P. Kot, M. Ismail, R. Drost, J. Siebrecht, H. Huang, and C. R. Ast, Electric control of spin transitions at the atomic scale, Nat. Commun. \textbf{14}, 6612 (2023).

\bibitem{ReinaGalvez2021} J. Reina-G\'alvez, N. Lorente, F. Delgado, and L. Arrachea, All-electric electron spin resonance studied by means of Floquet quantum master equations, Phys. Rev. B \textbf{104}, 245435 (2021).

\bibitem{Willke2018} P. Willke, W. Paul, F. D. Natterer, K. Yang, Y. Bae, T. Choi, J. Fern\'andez-Rossier, A. J. Heinrich, and C. P. Lutz, Probing quantum coherence in single-atom electron spin resonance, Sci. Adv. \textbf{4}, eaaq1543 (2018).

\bibitem{Seifert2020SA} T. S. Seifert, S. Kovarik, D. M. Juraschek, N. A. Spaldin, P. Gambardella, and S. Stepanow, Longitudinal and transverse electron paramagnetic resonance in a scanning tunneling microscope, Sci. Adv. \textbf{6}, eabc5511 (2020).

\bibitem{Seifert2020PRR} T. Seifert, S. Kovarik, C. Nistor, L. Persichetti, S. Stepanow, and P. Gambardella, Single-atom electron paramagnetic resonance in a scanning tunneling microscope driven by a radio-frequency antenna at 4 K, Phys. Rev. Research \textbf{2}, 013032 (2020).

\bibitem{Kovarik2024} S. Kovarik \textit{et al.}, Spin torque-driven electron paramagnetic resonance of a single spin in a pentacene molecule, Science \textbf{384}, 1368 (2024). 

\bibitem{Ferron2019} A. Ferr\'on, S. A. Rodr\'iguez, S. S. G\'omez, J. L. Lado, and J. Fern\'andez-Rossier, Single spin resonance driven by electric modulation of the $g$-factor anisotropy, Phys. Rev. Research \textbf{1}, 033185 (2019).

\bibitem{Balatsky2012} A. V. Balatsky, M. Nishijima, and Y. Manassen, Electron spin resonance-scanning tunneling microscopy, Adv. Phys. \textbf{61}, 117 (2012).

\bibitem{Sonntag2014} A. Sonntag \textit{et al.}, Electric-field-induced magnetic anisotropy in a nanomagnet investigated on the atomic scale, Phys. Rev. Lett. \textbf{112}, 017204 (2014). 

\bibitem{Korringa1950} J. Korringa, Nuclear magnetic relaxation and resonance line shift in metals, Physica \textbf{16}, 601 (1950). 

\bibitem{Heinrich2013} B. W. Heinrich, L. Braun, J. I. Pascual, and K. J. Franke, Protection of excited spin states by a superconducting energy gap, Nat. Phys. \textbf{9}, 765 (2013). 

\bibitem{Yu1965} L. Yu, Bound state in superconductors with paramagnetic impurities, Acta Phys. Sin. \textbf{21}, 75 (1965). 

\bibitem{Shiba1968} H. Shiba, Classical spins in superconductors, Prog. Theor. Phys. \textbf{40}, 435 (1968). 

\bibitem{Rusinov1969} A. I. Rusinov, Superconductivity near a paramagnetic impurity, JETP Lett. \textbf{9}, 85 (1969). 

\bibitem{Yazdani1997} A. Yazdani, B. A. Jones, C. P. Lutz, M. F. Crommie, and D. M. Eigler, Probing the local effects of magnetic impurities on superconductivity, Science \textbf{275}, 1767 (1997). 

\bibitem{Ji2008} S.-H. Ji \textit{et al.}, High-resolution scanning tunneling spectroscopy of magnetic impurity induced bound states in the superconducting gap of Pb thin films, Phys. Rev. Lett. \textbf{100}, 226801 (2008). 

\bibitem{Ruby2016} M. Ruby, Y. Peng, F. von Oppen, B. W. Heinrich, and K. J. Franke, Orbital picture of Yu-Shiba-Rusinov multiplets, Phys. Rev. Lett. \textbf{117}, 186801 (2016). 

\bibitem{Choi2017} D.-J. Choi \textit{et al.}, Mapping the orbital structure of impurity bound states in a superconductor, Nat. Commun. \textbf{8}, 15175 (2017). 

\bibitem{Heinrich2018} B. W. Heinrich, J. I. Pascual, and K. J. Franke, Single magnetic adsorbates on s-wave superconductors, Prog. Surf. Sci. \textbf{93}, 1 (2018). 

\bibitem{Franke2011} K. J. Franke, G. Schulze, and J. I. Pascual, Competition of superconducting phenomena and Kondo screening at the nanoscale, Science \textbf{332}, 940 (2011). 

\bibitem{Farinacci2018} L. Farinacci \textit{et al.}, Tuning the coupling of an individual magnetic impurity to a superconductor: quantum phase transition and transport, Phys. Rev. Lett. \textbf{121}, 196803 (2018). 

\bibitem{Kuester2021a} F. K\"uster, A. M. Montero, F. S. M. Guimar\~aes, S. Brinker, S. Lounis, S. S. P. Parkin, and P. Sessi, Long range and highly tunable interaction between local spins coupled to a superconducting condensate, Nat. Commun. \textbf{12}, 6722 (2021). 

\bibitem{Kuester2021b} F. K\"uster, A. M. Montero, F. S. M. Guimar\~aes, S. Brinker, S. Lounis, S. S. P. Parkin, and P. Sessi, Correlating Josephson supercurrents and Shiba states in quantum spins unconventionally coupled to superconductors, Nat. Commun. \textbf{12}, 1108 (2021).

\bibitem{Brinker2022} S. Brinker, F. K\"uster, S. S. P. Parkin, P. Sessi, and S. Lounis, Anomalous excitations of atomically crafted quantum magnets, Sci. Adv. \textbf{8}, eabi7291 (2022). 

\bibitem{Cornils2017} L. Cornils \textit{et al.}, Spin-resolved spectroscopy of the Yu-Shiba-Rusinov states of individual atoms, Phys. Rev. Lett. \textbf{119}, 197002 (2017). 

\bibitem{Beck2021} P. Beck \textit{et al.}, Spin-orbit coupling induced splitting of Yu-Shiba-Rusinov states in antiferromagnetic dimers, Nat. Commun. \textbf{12}, 2040 (2021). 

\bibitem{Schneider2021} L. Schneider \textit{et al.}, Topological Shiba bands in artificial spin chains on superconductors, Nat. Phys. \textbf{17}, 943 (2021). 

\bibitem{Wiesendanger2009} R. Wiesendanger, Spin mapping at the nanoscale and atomic scale, Rev. Mod. Phys. \textbf{81}, 1495 (2009).

\bibitem{Balatsky2006} A. V. Balatsky, I. Vekhter, and J.-X. Zhu, Impurity-induced states in conventional and unconventional superconductors, Rev. Mod. Phys. \textbf{78}, 373 (2006). 

\bibitem{Mishra2021} A. Mishra, P. Simon, T. Hyart, and M. Trif, Yu-Shiba-Rusinov qubit, PRX Quantum \textbf{2}, 040347 (2021). 

\bibitem{MicrowaveYSR2023} J. Siebrecht, H. Huang, P. Kot, R. Drost, C. Padurariu, B. Kubala, J. Ankerhold, J. C. Cuevas, and C. R. Ast, Microwave excitation of atomic scale superconducting bound states, Nat. Commun. \textbf{14}, 6794 (2023).

\bibitem{Ismail2025} M. Ismail, J. C. Cuevas, and C. R. Ast, Exploiting YSR states for driving spin transitions in a magnetic field, DPG Spring Meeting, Regensburg (2025). 

\bibitem{Etxeberria2026} M. Etxeberria-Etxaniz, A. Arnau, and A. Eiguren, Magnetism induced by periodically driven nonmagnetic impurities on surfaces with spin-orbit coupling, Phys. Rev. Lett. \textbf{136}, 176401 (2026).

\bibitem{Kambersky1976} V. Kambersk\'y, On ferromagnetic resonance damping in metals, Czech. J. Phys. B \textbf{26}, 1366 (1976). 

\bibitem{Kambersky2007} V. Kamberský, Spin-orbital Gilbert damping in common magnetic metals, Phys. Rev. B \textbf{76}, 134416 (2007). 

\bibitem{Guimaraes2019} F. S. M. Guimar\~aes, J. R. Suckert, J. Chico, J. Bouaziz, M. dos Santos Dias, and S. Lounis, Comparative study of methodologies to compute the intrinsic Gilbert damping: interrelations, validity and physical consequences, J. Phys.: Condens. Matter \textbf{31}, 255802 (2019). 

\bibitem{Lounis2010} S. Lounis, A. T. Costa, R. B. Muniz, and D. L. Mills, Dynamical magnetic excitations of nanostructures from first principles, Phys. Rev. Lett. \textbf{105}, 187205 (2010). 

\bibitem{Lounis2015} S. Lounis, M. dos Santos Dias, and B. Schweflinghaus, Transverse dynamical magnetic susceptibilities from regular static density functional theory: Evaluation of damping and $g$ shifts of spin excitations, Phys. Rev. B \textbf{91}, 104420 (2015).

\bibitem{Bouaziz2020} J. Bouaziz, F. S. Mendes Guimar\~aes, and S. Lounis, A new view on the origin of zero-bias anomalies of Co atoms atop noble metal surfaces, Nat. Commun. \textbf{11}, 6112 (2020). 

\bibitem{Friedrich2024} F. Friedrich, A. Odobesko, J. Bouaziz, S. Lounis, and M. Bode, Evidence for spinarons in Co adatoms, Nat. Phys. \textbf{20}, 28 (2024). 

\bibitem{Noei2023} N. Noei, R. Mozara, A. M. Montero, S. Brinker, N. Ide, F. S. M. Guimar\~aes, A. I. Lichtenstein, R. Berndt, S. Lounis, and A. Weismann, Manipulating the spin orientation of Co atoms using monatomic Cu chains, Nano Lett. \textbf{23}, 8988 (2023).

\bibitem{Duan2008} C.-G. Duan, J. P. Velev, R. F. Sabirianov, Z. Zhu, J. Chu, S. S. Jaswal, and E. Y. Tsymbal, Surface magnetoelectric effect in ferromagnetic metal films, Phys. Rev. Lett. \textbf{101}, 137201 (2008). 

\bibitem{Nakamura2009} K. Nakamura, R. Shimabukuro, Y. Fujiwara, T. Akiyama, T. Ito, and A. J. Freeman, Giant modification of the magnetocrystalline anisotropy in transition-metal monolayers by an external electric field, Phys. Rev. Lett. \textbf{102}, 187201 (2009).

\bibitem{Nakamura2009b} K. Nakamura, R. Shimabukuro, T. Akiyama, T. Ito, and A. J. Freeman, Origin of electric-field-induced modification of magnetocrystalline anisotropy at Fe(001) surfaces: Mechanism of dipole formation from first principles, Phys. Rev. B \textbf{80}, 172402 (2009).

\bibitem{Brovko2014} O. O. Brovko, P. Ruiz-D\'iaz, T. R. Dasa, and V. S. Stepanyuk, Controlling magnetism on metal surfaces with non-magnetic means: electric fields and surface charging, J. Phys.: Condens. Matter \textbf{26}, 093001 (2014).

\bibitem{Schweflinghaus2014} B. Schweflinghaus, M. dos Santos Dias, A. T. Costa, and S. Lounis, Renormalization of electron self-energies via their interaction with spin excitations: A first-principles investigation, Phys. Rev. B \textbf{89}, 235439 (2014). 

\bibitem{Khajetoorians2011} A. A. Khajetoorians, S. Lounis, B. Chilian, A. T. Costa, L. Zhou, D. L. Mills, J. Wiebe, and R. Wiesendanger, Itinerant nature of atom-magnetization excitation by tunneling electrons, Phys. Rev. Lett. \textbf{106}, 037205 (2011).

\bibitem{Khajetoorians2013} A. A. Khajetoorians, T. Schlenk, B. Schweflinghaus, M. dos Santos Dias, M. Steinbrecher, M. Bouhassoune, S. Lounis, J. Wiebe, and R. Wiesendanger, Spin excitations of individual Fe atoms on Pt(111): Impact of the site-dependent giant substrate polarization, Phys. Rev. Lett. \textbf{111}, 157204 (2013).

\bibitem{Schweflinghaus2016} B. Schweflinghaus, M. dos Santos Dias, and S. Lounis, Observing spin excitations in 3$d$ transition-metal adatoms on Pt(111) with inelastic scanning tunneling spectroscopy: A first-principles perspective, Phys. Rev. B \textbf{93}, 035451 (2016). 

\bibitem{SM} See Supplemental Material for the derivation of the identity and the selection rules, model details and parameters, robustness scans, the normal-state and Nambu damping calculations, convergence and gap-scaling checks, and the exact-diagonalization study.

\end{thebibliography}
\end{document}